\documentclass[9pt,conference]{IEEEtran}
\usepackage{cite}
\usepackage{amsmath,amssymb,amsfonts}
\usepackage{graphicx}
\usepackage{booktabs}
\usepackage{array}
\usepackage{textcomp}
\usepackage{xcolor}
\ifdefined\pdfsuppressptexinfo\pdfsuppressptexinfo=-1\fi  %
\usepackage{colortbl}
\usepackage{multirow}  %
\usepackage{capt-of}  %
\usepackage[ruled,vlined,linesnumbered]{algorithm2e}  %
\SetAlFnt{\footnotesize\linespread{0.85}\selectfont}  %
\SetVlineSkip{0.4ex}  %
\SetAlCapSkip{0.2ex}
\SetAlCapFnt{\small}
\SetAlCapNameFnt{\small}
\SetNlSty{textbf}{}{}
\SetArgSty{textup}

\SetCommentSty{algcommentfont}
\SetKwComment{tcp}{// }{}
\definecolor{linkblue}{rgb}{0,0,0.6}
\definecolor{algblue}{HTML}{0F70BD}  %

\usepackage[colorlinks=true,linkcolor=linkblue,citecolor=linkblue,urlcolor=linkblue]{hyperref}  %
\usepackage[capitalize,noabbrev]{cleveref}
\crefname{algocf}{Algorithm}{Algorithms}
\graphicspath{{figs/}}

\begin{document}

\title{CarrierKV: An Edge LLM Accelerator with Positional Carrier Decomposition for Low-Bit KV Cache}

\author{\IEEEauthorblockN{Junyi Luo\IEEEauthorrefmark{1}, Xinting Jiang\IEEEauthorrefmark{1}, Ruichen Qi\IEEEauthorrefmark{1}, Gregory Kielian\IEEEauthorrefmark{2}, Quan Cheng\IEEEauthorrefmark{1}, Mehdi Saligane\IEEEauthorrefmark{1}}
\IEEEauthorblockA{\IEEEauthorrefmark{1}Brown University\qquad \IEEEauthorrefmark{2}Google Research}
\IEEEauthorblockA{\{junyi\_luo, mehdi\_saligane\}@brown.edu}}

\maketitle

\begin{abstract}
Long-context decoding on edge devices is limited by DRAM, because every generated token reads the whole, growing KV cache.
4-bit block-scaled formats such as NVFP4 reduce this traffic but degrade accuracy: on Qwen3-4B, NVFP4 lowers 32K RULER (seven tasks) from 92.6 to 79.4, and a Hadamard rotation lowers it further to 57.7.
We observe that most of the key energy is a positional carrier, one per-head vector that rotary position embedding (RoPE) rotates with position.
In NVFP4, it sets the shared scale of most key blocks and lowers the signal-to-quantization-noise ratio of the content by about 5\,dB, nearly one bit, yet it depends only on that vector and the position.
We therefore introduce positional carrier decomposition, which removes the carrier before RoPE and stores only the content residual.
We design a KV path that rotates and quantizes keys only after this removal and computes each attention score directly on the 4-bit code.
We also design a positional phase engine that regenerates the carrier's score term for one key per cycle.
At the same cache size as NVFP4, CarrierKV{} raises 32K RULER on Qwen3-4B from 79.4 to 91.1 and lowers the 32K PG19 perplexity increase of NVFP4 by 1.8 to 5.9 times on four models, outperforming SageAttention and P3-LLM on all four.
Its loss also stays nearly flat as the cache grows, while that of NVFP4 rises.
In a 22\,nm FDSOI process at 600\,MHz, CarrierKV's hardware increases the area of the NVFP4 accelerator by 2.9\% and the energy of a 20K-token reasoning request on Qwen3-4B by 0.12\% at the same decode latency.
On Qwen3-4B at 32K, CarrierKV has a better RULER score than FP8 keys (91.1 against 89.3) and uses 13.7\% less energy per token.
The decomposition also lowers the 32K perplexity increase of 3-bit TurboQuant keys on Qwen3-4B from 69.7 to 0.76.
\end{abstract}

\begin{IEEEkeywords}
KV cache, NVFP4, RoPE, LLM accelerator.
\end{IEEEkeywords}

\section{Introduction}\label{sec:intro}

\begin{figure}[t]
\centering
\includegraphics[width=\columnwidth]{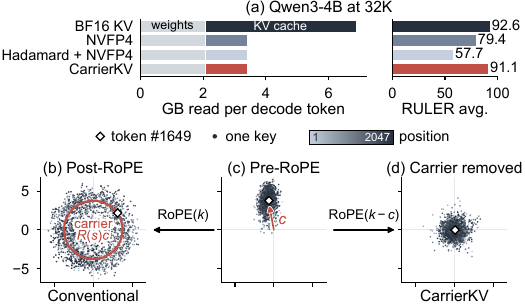}
\caption{(a) 4-bit KV caches reduce the bytes per decode token but degrade accuracy, which CarrierKV restores at the same cache size.
(b)--(d) Most key energy is a positional carrier: Qwen3-4B keys of one RoPE pair (layer 18) lie on the carrier's circle after RoPE, and a small residual remains once their mean $c$ is removed before RoPE.}
\label{fig:keypair}
\end{figure}

Long-context decoding on edge devices is limited by DRAM capacity and traffic.
At batch size one, every generated token reads all weights and the whole KV cache~\cite{qin2025bitlevel,hao2026nvllm}.
At 32K tokens, for example, the KV cache of Qwen3-4B~\cite{yang2025qwen3} takes 4.8\,GB in BF16, about 70\% of the bytes read per decode token with 4-bit weights (\cref{fig:keypair}(a)).
4-bit KV formats such as NVFP4~\cite{nvidia2025kv} reduce this cache by 3.6 times but degrade accuracy: on Qwen3-4B, NVFP4 KV lowers 32K RULER (seven tasks)~\cite{hsieh2024ruler} from 92.6 to 79.4.
We observe that most of the energy in the cached keys is a positional carrier, which can be computed from their positions, so a low-bit KV cache spends its bits and bandwidth on a component that need not be stored.

To reduce the accuracy loss of 4-bit keys, KV-cache accelerators refine how the cached keys are quantized: they store outliers~\cite{kim2025oaken} or a base key per group~\cite{li2025kvollm} separately, or adapt the number format of each group~\cite{hu2025mant,geng2026adana}.
Other designs change the keys before quantization: they subtract a fixed vector after rotary position embedding (RoPE)~\cite{sageattention2025,lee2026reqat,harmonia2026,tsai2026omnilut}, subtract a mean per chunk of tokens and store it~\cite{son2026nsnquant}, rotate them with a Hadamard transform~\cite{ashkboos2024quarot}, or quantize them before RoPE at the cost of RoPE on every read~\cite{hooper2024kvquant}.
None of them computes this carrier from the position with one fixed vector per KV head, and a Hadamard rotation even increases the loss: on Qwen3-4B, it lowers 32K RULER further, to 57.7.

Before RoPE, each key is a per-head constant vector plus a \emph{content residual}, and RoPE rotates the constant with position into the \emph{positional carrier} (\cref{fig:keypair}(b)--(d)).
This constant holds 61\% to 73\% of the key energy (\cref{sec:motivation}) and explains why the approaches above are insufficient.
It determines the scale of most NVFP4 blocks, so the content residual occupies only a few of the 4-bit levels and loses about 5\,dB of signal-to-quantization-noise ratio, nearly one bit; because it rotates with position, a fixed vector removes only its static part, and a Hadamard rotation distributes it across the blocks of the head.
Before RoPE, however, the carrier is one constant vector per KV head, so a single vector removes it entirely, with no metadata that grows with the context and no high-precision window, and the hardware can compute its effect from the position instead of storing it.

The removed carrier cannot simply be discarded: it contributes a query-dependent positional bias to every score, so the hardware must regenerate this term for every key and query head as fast as compressed keys stream from DRAM.
We therefore design CarrierKV, an NVFP4 edge LLM accelerator that stores only the compressed content residual and regenerates the carrier term on chip.

We evaluate four models from three families on RULER~\cite{hsieh2024ruler}, LongBench v1~\cite{bai2024longbench} and the perplexity (PPL) of 4K- and 32K-token PG19~\cite{rae2020compressive} windows.
At the same cache size as NVFP4, CarrierKV raises 32K RULER on Qwen3-4B from 79.4 to 91.1 (BF16 KV cache: 92.6) and lowers the 32K PPL increase of NVFP4 by 1.8 to 5.9 times, outperforming SageAttention~\cite{sageattention2025} and P3-LLM~\cite{p3llm2026} on all four models.
With 3-bit TurboQuant~\cite{zandieh2026turboquant} keys, the decomposition lowers the 32K PPL increase on Qwen3-4B from 69.7 to 0.76 (\cref{tab:tq}).
In a 22\,nm FDSOI process at 600\,MHz, CarrierKV's hardware increases the area of the NVFP4 accelerator by 2.9\% and the energy of a 20K-token reasoning request by 0.12\% on Qwen3-4B, with the same decode latency; against FP8 keys at a similar 32K PPL increase (FP8 keys: 0.173, CarrierKV: 0.199, NVFP4: 0.917), it reduces the energy per token at 32K by 13.7\% and is 1.16 times faster (\cref{sec:eval}).

This paper makes three contributions:
\begin{enumerate}
\item We show that most of the energy in the cached keys is a positional carrier that costs low-bit block formats nearly one bit, and introduce positional carrier decomposition, which removes it before RoPE with one vector per KV head and keeps the NVFP4 format and size.
\item We design a decompose-then-rotate KV path that applies the Hadamard rotation only after the carrier is removed and computes each attention score on the stored 4-bit code, so no cached key is reconstructed; only the query is rotated, once per step.
\item We design a positional phase engine that regenerates the carrier's score term on chip from the position, for one key per cycle, and integrate it with the KV path into an NVFP4 edge LLM accelerator.
\end{enumerate}

\Cref{sec:motivation} shows what the cached keys hold and why prior treatments are insufficient, \cref{sec:algorithm} presents the algorithm, \cref{sec:arch} the accelerator, and \cref{sec:eval} evaluates accuracy and hardware cost.

\section{Background and Motivation}\label{sec:motivation}

\begin{figure}[t]
\centering
\includegraphics[width=\columnwidth]{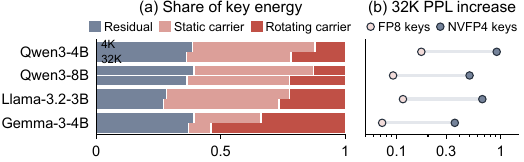}
\caption{(a) Key energy in the residual and the static and rotating carrier on PG19; (b) 32K PPL increase for FP8 or NVFP4 keys and NVFP4 values.}
\label{fig:energy}
\end{figure}

\textbf{KV cache in decode.}
During decode, each new token computes the attention score $q^\top k$ of its query $q$ with every cached key $k$, so every layer reads its whole KV cache per token~\cite{pope2023scaling}.
\textbf{RoPE.}
Rotary position embedding (RoPE)~\cite{su2024roformer3} splits a head of dimension $d$ into $d/2$ two-dimensional pairs and rotates pair $j$ at position $s$ by the angle $\theta_j s$.
The frequencies $\theta_j$ span several orders of magnitude, and we write $R(s)$ for the RoPE rotation of the whole head.
A vector shared by all keys before RoPE therefore rotates with position after RoPE, slowly in some pairs and rapidly in others (\cref{fig:keypair}(c) and (b)).
\textbf{Low-bit key formats.}
NVFP4 stores each element in 4 bits (E2M1) and shares one E4M3 scale among 16 elements~\cite{nvidia2025kv} (4.5 bits per element).
Because the scale follows the largest element of the block, one large value coarsens the grid of all 16 elements.
A randomized Hadamard rotation $H$ commonly flattens such outliers before 4-bit quantization~\cite{ashkboos2024quarot,liu2025spinquant}: it mixes all $d$ dimensions of the head, and the query is rotated by the same $H$.
\textbf{Centering.}
A common remedy for a component shared by all keys, which we call \emph{centering}, subtracts a fixed vector $m$ from the cached keys after RoPE, such as a mean over tokens~\cite{sageattention2025,lee2026reqat} or a per-channel offset~\cite{harmonia2026,tsai2026omnilut}.
Since $q^\top m$ is the same for every key, it cancels in the softmax.

\textbf{Observation 1: most of the energy in the cached keys is a positional carrier, and the keys cause most of the low-bit loss.}
We write a key before RoPE as $k(s) = c + r(s)$, where the vector $c \in \mathbb{R}^d$, the mean key of one KV head, is the carrier vector and $r(s)$ is the content residual; after RoPE, the cached key is $R(s)c + R(s)r(s)$, and we call $R(s)c$ the positional carrier (\cref{fig:keypair}(b)--(d)).
Earlier work observed the shared component~\cite{sun2024massive,gu2025attention,jin2025massive,son2026nsnquant} and corrected for parts of it~\cite{dualqk2026,iccad2026nvm}.
\Cref{fig:energy}(a) splits the key energy on PG19 into the residual and two parts of the carrier: a \emph{static} part, its mean over the context, and a \emph{rotating} part around this mean.
On four models, the carrier holds 61\% to 73\% of the key energy, measured as the squared norm of the keys.
Since RoPE and the Hadamard rotation preserve the norm, this share is the same before and after either rotation.
The rotating part of the carrier grows with context, on Qwen3-4B from 13\% at 4K to 22\% at 32K.
Values have no RoPE and hence no rotating carrier, and the keys account for the loss: with NVFP4 values, the 32K PPL increase is 5 to 6 times larger with NVFP4 keys than with FP8 keys (\cref{fig:energy}(b)).

\begin{figure}[t]
\centering
\includegraphics[width=\columnwidth]{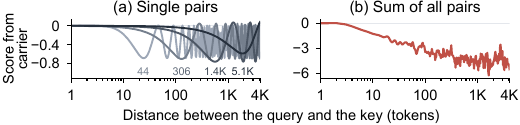}
\caption{Score from the carrier relative to the nearest key, for the mean query of one Qwen3-4B head: (a) four pairs, periods in tokens; (b) all pairs.}
\label{fig:pulse}
\end{figure}

\begin{table}[t]
\centering
\caption{Approaches to the shared key component.}
\label{tab:prior}
\fontsize{6.2}{7.4}\selectfont
\setlength{\tabcolsep}{1pt}
\renewcommand{\arraystretch}{1.05}
\setlength{\aboverulesep}{0.2ex}\setlength{\belowrulesep}{0.2ex}
\newcommand{\hd}[1]{{\fontsize{6.5}{7.5}\selectfont\textbf{#1}}}
\begin{tabular*}{\columnwidth}{@{\extracolsep{\fill}}lllll@{}}
\toprule[0.8pt]
\hd{Approach} & \hd{Key change} & \hd{Extra storage} & \hd{Work per read} & \hd{Rotating carrier} \\
\midrule
SageAttention~\cite{sageattention2025} & $-$vector post-RoPE & vector per KV head & none & left in keys \\
\rowcolor{black!7}Hadamard~\cite{ashkboos2024quarot} & apply $H$ & none & apply $H$ to query & across blocks \\
P3-LLM~\cite{p3llm2026} & $\div$ channel scale & scale per channel & $\times$ scale on query & left in keys \\
\rowcolor{black!7}NSNQuant~\cite{son2026nsnquant} & \begin{tabular}[c]{@{}l@{}}$-$mean per\\64 tokens\end{tabular} & \begin{tabular}[c]{@{}l@{}}mean per 64 tokens,\\recent keys in FP16\end{tabular} & RoPE on means & \begin{tabular}[c]{@{}l@{}}removed per\\64 tokens\end{tabular} \\
SpectrumQuant~\cite{spectrumquant2026} & DCT per 128 tokens & \begin{tabular}[c]{@{}l@{}}top DCT terms,\\recent keys in FP16\end{tabular} & inverse DCT & in DCT terms \\
\rowcolor{black!7}KVQuant~\cite{hooper2024kvquant} & quantize pre-RoPE & none & RoPE per key & never stored \\
\midrule
CarrierKV & $-c$ pre-RoPE & $c$ per KV head & $q^\top R(s)c$ per key & regenerated \\
\bottomrule[0.8pt]
\end{tabular*}
\end{table}

\textbf{Observation 2: centering and the Hadamard rotation rest on assumptions that the carrier breaks.}
Centering assumes that the shared vector is the same in every cached key, but after RoPE it rotates with position, so, like the removal of a DC offset from a signal, a fixed $m$ removes at most the static part of the carrier.
On NVFP4 keys, centering with the mean of the 512 prompt tokens after the first lowers the 32K PPL increase on Qwen3-4B only from 0.917 to 0.688, and its loss grows with the cache, from 0.48 in the first 4K to 0.85 in the last 4K.
The Hadamard rotation assumes that no component dominates the head, but in the first layers of Qwen3 the carrier does, and the rotation distributes it across the blocks: on Qwen3-4B, the 32K PPL increase grows from 0.917 with NVFP4 to 3.50, and 32K RULER decreases from 79.4 to 57.7.
The carrier must therefore be removed before RoPE, where it is still constant, and before the Hadamard rotation.

\textbf{Observation 3: the carrier is a positional bias that the model relies on.}
The carrier term $q^\top R(s)c$ of the score depends only on the query and the position, and each pair contributes one sinusoid in the distance between the query and the key.
\Cref{fig:pulse} shows that in a trained model, $c$ provides the preference for near keys that RoPE was designed for~\cite{su2024roformer3}: fast pairs separate the nearest keys, and slow pairs separate keys thousands of tokens apart.
Across the 1152 query heads of Qwen3-4B, the carrier term varies over distance by a median of 2.8 times the spread of the content term $q^\top R(s)r(s)$, and 95\% of the heads favor keys 1 to 8 tokens away by more than 0.5 in score, in line with earlier reports~\cite{barbero2024round,triattention2026}.
Subtracting $c$ from the stored keys removes this bias at every distance, so the read path must restore the term for every pair, including the slowest ones: if the nine slowest pairs are subtracted but not regenerated, the 32K PPL increase on Qwen3-4B grows from 0.199 to 0.456.

\textbf{Prior handling of the shared component.}
\Cref{tab:prior} lists the prior approaches; Observation~2 covers centering and the Hadamard rotation.
NSNQuant~\cite{son2026nsnquant} subtracts a channel mean for every 64 tokens inside its vector quantizer, stores these means, and keeps the newest, unfinished 64 tokens in FP16.
SpectrumQuant~\cite{spectrumquant2026} quantizes a discrete cosine transform (DCT) over every 128 tokens, keeps its largest coefficients in FP16 and applies an inverse DCT to the scores on every read.
KVQuant~\cite{hooper2024kvquant} stores keys before RoPE, where the shared component does not rotate, but every decode step must then decode every cached key and apply RoPE to it before the dot product.
P3-LLM~\cite{p3llm2026} does so only for Llama-1 and~2, which moves the score computation off PIM; for newer models it scales each key channel after RoPE and leaves the carrier in the keys.
\Cref{sec:algorithm} answers Observations~1 and~2, and \cref{sec:arch} answers Observation~3.

\section{CarrierKV Algorithm}\label{sec:algorithm}

\begin{algorithm}[b]
\caption{CarrierKV for one KV head. $R(s)$ is RoPE at position $s$, $H$ the Hadamard matrix, $\theta_j$ the RoPE frequency of pair $j$, and $q_{h,j}$, $c_j$ are pair $j$ of $q_h$ and $c$.}\label{alg:carrierkv}
\SetCommentSty{algcmt}
\KwIn{keys $k(s)$ before RoPE, values $v(s)$, prompt length $L_p$;\\\hphantom{\textbf{Input:}\ }queries $q_h$ after RoPE of its query heads $h$}
\KwOut{attention output $o_h$ of every query head and decode step}
$c \gets$ mean of $k(s)$ over $1 \le s \le \min(512, L_p - 1)$\label{alg:line:c}\tcp*[r]{prefill}
\tcp{\textbf{Write path:} decompose, then rotate}
\For{each token $s$, prompt and generated}{
  $r(s) \gets k(s) - c$\label{alg:line:r}\tcp*[r]{decomposition}
  $\hat k(s) \gets \texttt{NVFP4}(H R(s)\, r(s))$\label{alg:line:cache}\tcp*[r]{RoPE, then $H$}
  $\hat v(s) \gets \texttt{NVFP4}(v(s))$\;
}
\tcp{\textbf{Read path:} score from the code and the position}
\For{each decode step}{
  \For{each query head $h$, with query $q_h$}{
    $\hat q_h \gets H q_h$\label{alg:line:qhat}\tcp*[r]{once per step}
    $a_{h,j} \gets q_{h,j} \cdot c_j$, \ $b_{h,j} \gets q_{h,j} \times c_j$ for every pair $j$\label{alg:line:coef}\;
  }
  $\phi_j \gets 0$ for every pair $j$\tcp*[r]{phase engine}
  \For{each cached position $s = 0, 1, \ldots$}{
    look up $\cos\phi_j$ and $\sin\phi_j$ for every pair $j$\label{alg:line:cossin}\tcp*[r]{shared}
    \For{each query head $h$}{
      $\mathrm{score}_h(s) \gets \hat q_h^\top \hat k(s) + \sum_j (a_{h,j}\cos\phi_j - b_{h,j}\sin\phi_j)$\label{alg:line:score}\;
    }
    $\phi_j \gets \phi_j + \theta_j$ for every pair $j$\label{alg:line:phase}\tcp*[r]{adds only}
  }
  \For{each query head $h$}{
    $o_h \gets \sum_s p_h(s)\, \hat v(s)$ with $p_h = \texttt{softmax}(\mathrm{score}_h)$\;
  }
}
\end{algorithm}

\begin{figure}[t]
\centering
\includegraphics[width=\columnwidth]{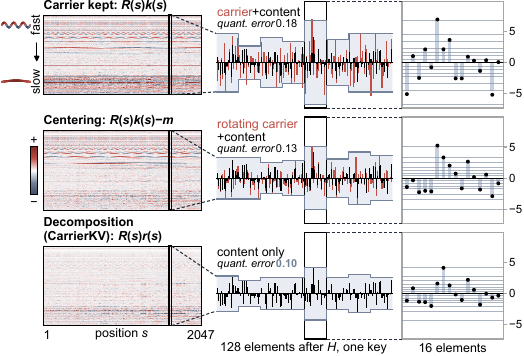}
\caption{One Qwen3-4B KV head under three treatments: the cached keys over positions (left), the key at position 1649 after $H$ with its carrier part in red and each NVFP4 block's range (middle), and its block with the largest scale on the E2M1 grid (right).}
\label{fig:blocks}
\end{figure}

CarrierKV caches each key without its positional carrier and computes the carrier's share of the score from the position (\cref{alg:carrierkv}), with three advantages over prior treatments of the keys.
First, the NVFP4 block scales are set by the content rather than by the carrier, which raises the signal-to-quantization-noise ratio of the content by about 5\,dB, nearly one bit; centering after RoPE~\cite{sageattention2025} removes only the non-rotating part of the carrier (\cref{fig:blocks}), per-channel scaling~\cite{p3llm2026} rescales the carrier without removing it, and a Hadamard rotation~\cite{ashkboos2024quarot} distributes it across the blocks.
Second, the cache retains the NVFP4 format and footprint: $c$ adds 0.0002 bits per key element at 32K, whereas the per-chunk mean subtraction of NSNQuant~\cite{son2026nsnquant} adds 0.23 bits of metadata per element, three orders of magnitude more, and a high-precision (FP16) window of recent keys.
Third, the content term is computed on the 4-bit codes and the carrier term without memory access per key, whereas keys cached before RoPE~\cite{hooper2024kvquant} must be decoded and rotated on every read.

\textbf{Positional carrier decomposition.}
For each KV head, CarrierKV estimates the carrier vector $c \in \mathbb{R}^d$, the mean key of the head, with $d$ the head dimension, during prefill: it averages the keys $k(s)$ before RoPE over up to 512 prompt tokens, skipping the first, which is an attention sink~\cite{xiao2024streamingllm}, and the keys of these tokens are buffered until $c$ is known.
With only 16 tokens, the 32K PPL increase of CarrierKV grows by at most 0.10 and stays far below that of NVFP4 on all four models.
It then caches only the content residual $r(s) = k(s) - c$ (\cref{alg:carrierkv}, lines~\ref{alg:line:c} and~\ref{alg:line:r}); since RoPE is linear, this removes the whole carrier $R(s)c$, static and rotating parts, at every position $s$.

\textbf{The carrier in NVFP4 blocks.}
When the largest element of an NVFP4 block is dominated by the carrier, the eight E2M1 magnitudes span the range of the carrier, and the much smaller content occupies only a few of them.
The three rows of \cref{fig:blocks} keep the whole carrier, remove its static part by centering (\cref{sec:motivation}) and remove all of it with CarrierKV.
For one Qwen3-4B key after $H$ (middle), the carrier part (red) sets the block ranges, so the quantization error relative to its content residual $R(s)r(s)$ falls from 0.18 to 0.13 and 0.10.
Over every layer of the four models on four 32K PG19 windows, the carrier part outweighs the content residual in 57\% to 72\% of the NVFP4 blocks after $H$ and makes their quantization step 1.7 to 1.9 times coarser than the content alone requires.
The signal-to-quantization-noise ratio of the content residual is 20.4\,dB with CarrierKV on all four models and 14.5 to 15.7\,dB with the carrier kept, a loss of 0.8 to 1.0 bits at about 6\,dB per bit.

\textbf{Score from the code and the position.}
Let $\hat k(s) = HR(s)r(s)$ be the key that CarrierKV quantizes to NVFP4 and caches, and let $q$ be a query after RoPE, scaled by $1/\sqrt{d}$.
With $k(s) = c + r(s)$, the score of $q$ with the key at position $s$ splits into a content term, from $r(s)$, and a carrier term, from $c$, in Eq.~\eqref{eq:split}.
In the content term, $H^\top H = I$ gives $q^\top R(s)r(s) = (Hq)^\top HR(s)r(s) = \hat q^\top \hat k(s)$ with $\hat q = Hq$.
In the carrier term, $R(s)$ rotates pair~$j$ of $c$, $c_j = (c_{j,1}, c_{j,2})$, by the angle $s\theta_j$; expanding the product with pair~$j$ of $q$, $q_j = (q_{j,1}, q_{j,2})$, gives $a_j = q_{j,1}c_{j,1} + q_{j,2}c_{j,2}$ and $b_j = q_{j,1}c_{j,2} - q_{j,2}c_{j,1}$, the dot and the cross product of $q_j$ and $c_j$:
\begin{align}
q^\top R(s)k(s) &= q^\top R(s)r(s) + q^\top R(s)c \nonumber\\
&= \hat q^\top \hat k(s) + \sum_{j=1}^{d/2} q_j^\top \begin{bmatrix} \cos(s\theta_j) & -\sin(s\theta_j) \\ \sin(s\theta_j) & \cos(s\theta_j) \end{bmatrix} c_j \nonumber\\
&= \hat q^\top \hat k(s) + \sum_{j=1}^{d/2} \bigl[a_j\cos(s\theta_j) - b_j\sin(s\theta_j)\bigr]. \label{eq:split}
\end{align}
Before quantization, Eq.~\eqref{eq:split} is exact for any fixed $c$: the estimate only sets how small the residual, and thus its quantization error, is.
CarrierKV therefore fixes $c$ after prefill and encodes each generated key as soon as it is produced.
Since each cached element is $S s_b e_i$, with the E2M1 element $e_i$, the E4M3 scale $s_b$ of block $b$ and the FP32 tensor scale $S$, the content term is $S \sum_b s_b \sum_{i \in b} \hat q_i e_i$ (line~\ref{alg:line:score}): each 4-bit code is multiplied by a query element and each scale is applied once per block, so no cached key is decoded to high precision.

The coefficients $a_j$ and $b_j$ of the carrier term are computed from $q$ and $c$ once per step and query head (line~\ref{alg:line:coef}), and its $\cos(s\theta_j)$ and $\sin(s\theta_j)$, which depend only on the position, are shared by all query heads of the KV head (line~\ref{alg:line:cossin}); each pair thus costs two multiply-accumulates per key and query head.
As the keys are read in order of position, the phase $s\theta_j$ advances by $\theta_j$ per key with additions only (line~\ref{alg:line:phase}), in the phase engine of \cref{sec:arch}; CarrierKV adds 0.12\% energy to a reasoning request over NVFP4 on Qwen3-4B (\cref{sec:eval}).

\begin{figure*}[t]
\centering
\includegraphics[width=\textwidth]{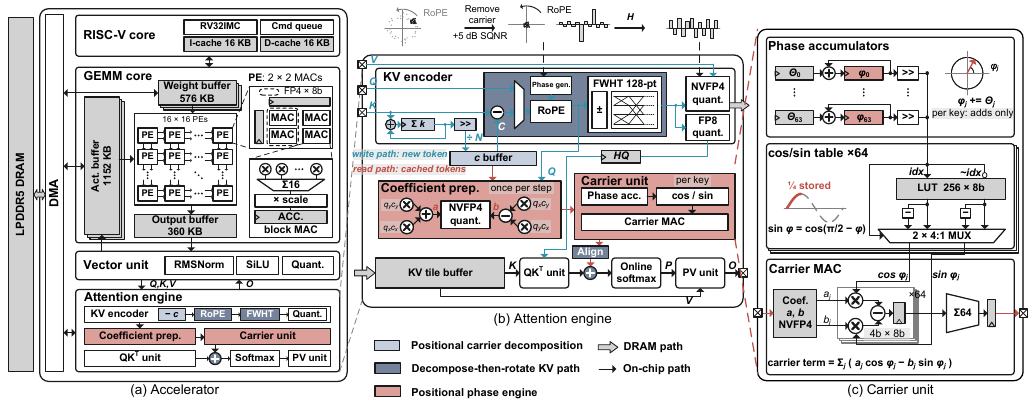}
\caption{CarrierKV accelerator: (a) the accelerator, (b) its attention engine and (c) the carrier unit.}
\label{fig:arch}
\end{figure*} 

\textbf{Decompose, then rotate.}
The write path subtracts $c$ before RoPE, where $c$ is constant, and applies $H$ after RoPE, so that $H$ can be applied to the query instead (line~\ref{alg:line:cache}).
Keys cached before RoPE, as in KVQuant~\cite{hooper2024kvquant}, need RoPE on every read: each cached key is decoded to high precision, rotated with four multiplications per pair and multiplied with each query in high precision, and with $H$ it would also have to be transformed back by $H^\top$ first.
CarrierKV instead rotates only the query, once per step, and adds two narrow multiply-accumulates per pair, key and query head for the carrier term, far less computation (\cref{sec:eval}).
Removing the carrier before $H$ also keeps the rotation from distributing a dominant component across the blocks (\cref{tab:ablation}).

\textbf{Ablation.}
\Cref{tab:ablation} compares CarrierKV with plain NVFP4 keys, centering, the Hadamard rotation and decomposition alone, in the deployment of \cref{sec:eval}; its last column stores the keys of CarrierKV at 3.5 bits per element, with INT3 elements (sign and magnitude 0 to 3) under the scales of NVFP4:
\par\smallskip
\noindent\begin{minipage}{\columnwidth}
\captionof{table}{Block-scaled keys: 32K PG19 PPL increase.}\label{tab:ablation}
\noindent{\fontsize{7.5}{9}\selectfont\setlength{\tabcolsep}{1.2pt}\renewcommand{\arraystretch}{0.88}%
\setlength{\aboverulesep}{0.2ex}\setlength{\belowrulesep}{0.2ex}%
\begin{tabular*}{\columnwidth}{@{\extracolsep{\fill}}lcccccc@{}}
\toprule[0.5pt]
Model & NVFP4 & Centering & Hadamard & Decomposition & \multicolumn{2}{c}{CarrierKV} \\
Cached key & $R(s)k$ & $R(s)k - m$ & $HR(s)k$ & $R(s)(k - c)$ & \multicolumn{2}{c}{$HR(s)(k - c)$} \\
\cmidrule(l){6-7}
Bits per element & 4.5 & 4.5 & 4.5 & 4.5 & 4.5 & 3.5 \\
\midrule[0.4pt]
Qwen3-4B & +0.917 & +0.688 & +3.50 & +0.204 & \textbf{+0.199} & +0.766 \\
Qwen3-8B & +0.503 & +0.420 & +1.51 & +0.103 & \textbf{+0.086} & +0.346 \\
Llama-3.2-3B & +0.667 & +0.831 & +0.451 & +0.564 & \textbf{+0.363} & +1.35 \\
Gemma-3-4B & +0.363 & +0.508 & +0.437 & +0.119 & \textbf{+0.102} & +0.333 \\
\bottomrule[0.5pt]
\end{tabular*}}
\end{minipage}
\par\smallskip

Centering lowers the 32K PPL increase on the Qwen3 models but raises it on Llama-3.2-3B and Gemma-3-4B, and the rotation alone raises it on the Qwen3 models and Gemma-3-4B.
On Gemma-3-4B, centering also raises the share of NVFP4 blocks that the carrier outweighs to 73\%, while it lowers this share to 36\% to 41\% on the other three models.
Decomposition alone lowers the PPL increase of NVFP4 on all four models, and the rotation after it lowers it further, most on Llama-3.2-3B (0.564 to 0.363).
At 3.5 bits per key element, CarrierKV still stays below 4.5-bit NVFP4 on Qwen3-4B, Qwen3-8B and Gemma-3-4B (0.766, 0.346 and 0.333 against 0.917, 0.503 and 0.363), but not on Llama-3.2-3B (1.35 against 0.667).

\textbf{Independence of the number format.}
The decomposition does not depend on the number format.
Applied in software to \textbf{TurboQuant}~\cite{zandieh2026turboquant} (MSE variant: Hadamard rotation, Lloyd-Max codebook, FP16 norm) with 3-bit keys and 4-bit values, in the same deployment, it gives the 32K PPL increase on PG19 in \cref{tab:tq}:
\par\smallskip
\noindent\begin{minipage}{\columnwidth}
\captionof{table}{3-bit TurboQuant keys: 32K PG19 PPL increase.}\label{tab:tq}
{\centering\footnotesize\setlength{\tabcolsep}{5pt}\renewcommand{\arraystretch}{0.9}%
\setlength{\aboverulesep}{0.2ex}\setlength{\belowrulesep}{0.2ex}%
\begin{tabular}{@{}lrrr@{}}
\toprule[0.5pt]
Model & Untreated & Centering & Decomposition \\
\midrule[0.4pt]
Qwen3-4B & +69.7 & +2.94 & \textbf{+0.762} \\
Qwen3-8B & +20.1 & +1.56 & \textbf{+0.329} \\
Llama-3.2-3B & \textbf{+1.89} & +2.15 & +1.93 \\
Gemma-3-4B & +1.21 & +1.48 & \textbf{+0.335} \\
\bottomrule[0.5pt]
\end{tabular}\par}
\end{minipage}
\par\smallskip

On Qwen3-4B, the decomposition lowers the PPL increase from 69.7 to 0.76.
It is lower than centering on all four models and lower than untreated keys on three; on Llama-3.2-3B it brings no gain (1.93 against 1.89 untreated).
With 2-bit keys, the decomposition gives the lowest PPL increase of the three on all four models, Llama-3.2-3B included, but 2-bit keys remain lossy, with increases of 1.3 to 12.6.

\section{CarrierKV Architecture}\label{sec:arch}

\textbf{Overview.}
\Cref{fig:arch}(a) shows the accelerator.
A $32 \times 32$ array of NVFP4 block multiply-accumulate units (MACs), arranged as $16 \times 16$ processing elements (PEs) of $2 \times 2$ MACs, computes the linear layers in W4A4, a vector unit computes RMSNorm, SiLU and the NVFP4 quantization of the GEMM inputs, and the attention engine computes the attention core (QK$^\top$, softmax and PV); a RISC-V core with 16\,KB instruction and data caches issues their tile commands.
The attention engine is designed for decode, where the query of each new token meets every cached key, one per cycle, as the keys stream from DRAM.
For the prompt, all queries meet all keys at once, so the GEMM core computes this attention as a matrix product, with a 16-row softmax and with the query, the softmax output $P$ and the values requantized to 8-bit blocks.
Its FP4 $\times$ 8-bit mode splits each 8-bit operand into two 4-bit digits for two MACs of a PE and combines their results by a shift and an add; QK$^\top$ uses this mode with 8-bit queries and the NVFP4 keys.
Weights and the NVFP4 KV cache reside in LPDDR5-6400 x64 DRAM, and about 2.1\,MB of SRAM buffers weights, activations and KV tiles.
The attention engine reads one key and one value of 128 elements per cycle, more than the DRAM delivers, so decode remains bound by DRAM bandwidth; a head of 256 dimensions takes two cycles per key, as does its DRAM read.
Since the carrier can be computed from the position, the accelerator regenerates it on chip instead of storing it.
CarrierKV adds three parts to an NVFP4 accelerator, one per contribution and color in \cref{fig:arch}.

\textbf{Positional carrier decomposition.}
During prefill, an accumulator in the KV encoder sums the keys before RoPE of the $N$ prompt tokens after the first, with $N$ the largest power of two up to $\min(512, L_p - 1)$, and a shift divides the sum by $N$ (\cref{alg:carrierkv}, line~\ref{alg:line:c}).
The result is quantized to 8-bit mantissas with one power-of-two exponent per KV head, stored in DRAM next to the KV cache and prefetched once per layer into the $c$ buffer, which serves both the subtraction and the coefficient preparation.
On Qwen3-4B, this prefetch reads about 36\,KB per decode step over its 36 layers and 8 KV heads, about 0.003\% of the keys and values that a step reads at 32K.
For every new token, a subtractor aligns $c$ to the exponent of the key and removes it before RoPE (line~\ref{alg:line:r}).

\textbf{Decompose-then-rotate KV path.}
After the subtraction, the key passes through RoPE, whose cos and sin come from its own phase generator with 32-bit phases and 10-bit values, and through a 128-point fast Walsh--Hadamard transform (FWHT) after random sign flips; the NVFP4 quantizer then stores it with one E4M3 scale per 16 elements (line~\ref{alg:line:cache}).
Values are quantized to NVFP4 directly.
Once per decode step, the same RoPE and FWHT rotate each query into $\hat q$ (line~\ref{alg:line:qhat}), so the read path needs no rotation unit.
The attention engine serves one KV head and its $G$ query heads at a time and streams one key per cycle from the KV tile buffer.
The QK$^\top$ unit computes the content term directly on the stored codes and scales with the 8-bit query $\hat q$ (line~\ref{alg:line:score}).
An adder then aligns the carrier term to this score and adds it before the online softmax~\cite{stevens2021softermax}.

\textbf{Positional phase engine.}
RoPE and the carrier term both need $\cos(s\theta_j)$ and $\sin(s\theta_j)$ for every pair, and a table over a 128K context would require tens of megabytes.
At the start of each step, the coefficient preparation computes $a_{h,j}$ and $b_{h,j}$ from $c$ and each query before the FWHT, with four multiplications per pair (line~\ref{alg:line:coef}).
It then quantizes them to NVFP4, with one E4M3 scale per 8 pairs.
The carrier unit then works like the numerically controlled oscillator of a digital receiver (\cref{fig:arch}(c)).
Per pair, a 28-bit phase accumulator, counted in turns, is cleared before the first key of each KV head, adds $\theta_j/2\pi$ per key and wraps around at a full turn (line~\ref{alg:line:phase}).
Its top 2 bits select the quadrant, and its next 8 bits address a table that stores the cosine over a quarter turn in 256 entries of 8 bits.
The table is read at this index and at its complement, which give the cosine and the sine within the quadrant, and the quadrant bits swap the two and set their signs.
The rounded step adds at most 0.0015\,rad by position 128K, a quarter of one table step.
One set of 64 cos and sin values per key serves all $G$ query heads (line~\ref{alg:line:cossin}).
A carrier MAC of $G \times 64 \times 2$ multipliers of 4 by 8 bits forms the carrier term of each query head from them, one key per cycle, so the carrier term adds no latency per key.
For the prompt, the GEMM core computes the carrier term in the same FP4 $\times$ 8-bit mode as QK$^\top$.

\section{Evaluation}\label{sec:eval}

\begin{table*}[t]
\centering
\caption{Accuracy on four models: 32K RULER, LongBench and PPL increase over the BF16 KV cache. Sage: SageAttention; P3: P3-LLM.}
\label{tab:accuracy}
\fontsize{7.5}{7.9}\selectfont
\setlength{\tabcolsep}{1pt}
\renewcommand{\arraystretch}{0.88}
\setlength{\aboverulesep}{0.2ex}\setlength{\belowrulesep}{0.2ex}
\begin{tabular*}{\textwidth}{@{\extracolsep{\fill}}clcccccccccccccccccccc@{}}
\toprule[0.8pt]
 & & \multicolumn{5}{c}{\textbf{Qwen3-4B}} & \multicolumn{5}{c}{\textbf{Qwen3-8B}} & \multicolumn{5}{c}{\textbf{Llama-3.2-3B}} & \multicolumn{5}{c}{\textbf{Gemma-3-4B}} \\
\cmidrule(lr){3-7}\cmidrule(lr){8-12}\cmidrule(lr){13-17}\cmidrule(lr){18-22}
 & \textbf{Task} & BF16 & NVFP4 & Sage & P3 & CarrierKV & BF16 & NVFP4 & Sage & P3 & CarrierKV & BF16 & NVFP4 & Sage & P3 & CarrierKV & BF16 & NVFP4 & Sage & P3 & CarrierKV \\
\midrule
\multirow{8}{*}{\rotatebox[origin=c]{90}{\textbf{32K RULER}}} & MK1 & 98.0 & 94.0 & 95.0 & 94.0 & \textbf{99.0} & 98.0 & \textbf{99.0} & 97.0 & 97.0 & 97.0 & 89.0 & 77.0 & 71.0 & \textbf{87.0} & 85.0 & 60.0 & 49.0 & 51.0 & \textbf{57.0} & 56.0 \\
 & MK2 & 82.0 & 58.0 & 77.0 & 44.0 & \textbf{85.0} & 96.0 & 85.0 & 94.0 & 91.0 & \textbf{94.0} & 66.0 & 45.0 & 38.0 & 53.0 & \textbf{59.0} & 12.0 & 12.0 & 11.0 & 13.0 & \textbf{13.0} \\
 & MK3 & 87.0 & 43.0 & 69.0 & 31.0 & \textbf{79.0} & 98.0 & 91.0 & 85.0 & 85.0 & \textbf{96.0} & 17.0 & 2.0 & 5.0 & 9.0 & \textbf{13.0} & 3.0 & 3.0 & 2.0 & 3.0 & \textbf{4.0} \\
 & MQ & 98.2 & 95.8 & 95.8 & 93.8 & \textbf{96.5} & 98.2 & 96.5 & 98.5 & 97.2 & \textbf{98.8} & 86.5 & 81.8 & 80.0 & 81.0 & \textbf{87.8} & 72.0 & 69.0 & 66.2 & \textbf{70.8} & 70.2 \\
 & MV & 96.8 & 92.5 & 92.0 & 92.2 & \textbf{96.5} & 97.5 & 96.5 & 97.2 & 96.0 & \textbf{97.2} & 85.0 & 84.2 & 80.0 & 84.5 & \textbf{84.5} & 55.2 & 53.8 & 53.2 & \textbf{55.2} & 54.8 \\
 & S3 & 100.0 & 99.0 & \textbf{100.0} & 99.0 & 99.0 & 100.0 & 100.0 & 100.0 & 100.0 & \textbf{100.0} & 96.0 & 93.0 & 87.0 & 88.0 & \textbf{100.0} & 100.0 & 100.0 & 100.0 & 100.0 & \textbf{100.0} \\
 & VT & 86.4 & 73.2 & 79.2 & 71.6 & \textbf{82.6} & 99.6 & 97.4 & 97.4 & 97.4 & \textbf{98.6} & 69.6 & 63.4 & 61.8 & \textbf{69.4} & 65.0 & 56.4 & \textbf{58.0} & 53.4 & 50.4 & 54.2 \\
 & Avg. & 92.6 & 79.4 & 86.9 & 75.1 & \textbf{91.1} & 98.2 & 95.1 & 95.6 & 94.8 & \textbf{97.4} & 72.7 & 63.8 & 60.4 & 67.4 & \textbf{70.6} & 51.2 & 49.2 & 48.1 & 49.9 & \textbf{50.3} \\
\midrule
\multirow{7}{*}{\rotatebox[origin=c]{90}{\textbf{LongBench}}} & SDoc & 42.9 & 40.5 & \textbf{42.9} & 41.0 & 40.9 & 44.7 & \textbf{44.4} & 43.3 & 43.8 & 44.0 & 41.8 & 40.2 & 40.9 & 40.2 & \textbf{41.3} & 41.6 & \textbf{40.7} & 40.5 & 39.9 & 40.4 \\
 & MDoc & 36.1 & 32.2 & \textbf{32.8} & 30.8 & 32.7 & 38.4 & 36.8 & 38.3 & 37.8 & \textbf{38.8} & 35.6 & 34.4 & 33.1 & \textbf{34.8} & 33.3 & 34.6 & 33.5 & 33.2 & 33.1 & \textbf{34.0} \\
 & Summ & 24.3 & 24.0 & \textbf{24.3} & 24.0 & 24.0 & 24.0 & 24.0 & 24.0 & 23.9 & \textbf{24.1} & 24.9 & \textbf{25.2} & 24.9 & 24.9 & 24.9 & 20.8 & 20.7 & 20.9 & 20.8 & \textbf{21.1} \\
 & Few & 61.6 & 59.4 & 59.7 & 59.8 & \textbf{60.3} & 62.5 & 62.8 & 62.7 & 62.8 & \textbf{62.9} & 59.1 & 56.5 & 57.3 & 57.8 & \textbf{58.0} & 53.9 & 52.1 & 52.6 & 51.1 & \textbf{53.1} \\
 & Syn & 47.3 & 49.6 & 45.5 & \textbf{50.0} & 49.7 & 56.6 & \textbf{59.3} & 56.8 & 53.4 & 57.1 & 37.4 & \textbf{33.0} & 29.2 & 32.9 & 31.9 & 30.5 & 25.0 & 25.0 & 25.5 & \textbf{27.5} \\
 & Code & 57.7 & 55.7 & 55.0 & 55.2 & \textbf{56.2} & 61.4 & 59.5 & 60.8 & 59.1 & \textbf{60.8} & 47.8 & 46.3 & 41.9 & 44.6 & \textbf{47.4} & 56.7 & 55.0 & 54.4 & \textbf{55.1} & 54.5 \\
 & Avg. & 43.64 & 42.11 & 42.16 & 42.03 & \textbf{42.53} & 46.23 & 46.12 & 45.95 & 45.31 & \textbf{46.30} & 40.64 & 38.89 & 37.91 & 38.97 & \textbf{39.08} & 38.51 & 36.80 & 36.82 & 36.47 & \textbf{37.42} \\
\midrule
\multirow{2}{*}[1pt]{\rotatebox[origin=c]{90}{\textbf{{\fontsize{4}{4}\selectfont$\Delta$}{\fontsize{5.5}{5.5}\selectfont PPL}}}} & 4K & -- & 0.861 & 0.469 & 0.469 & \textbf{0.262} & -- & 0.426 & 0.282 & 0.300 & \textbf{0.106} & -- & 0.571 & 0.663 & \textbf{0.348} & 0.349 & -- & 0.381 & 0.492 & 0.492 & \textbf{0.112} \\
 & 32K & -- & 0.917 & 0.688 & 0.968 & \textbf{0.199} & -- & 0.503 & 0.420 & 0.520 & \textbf{0.086} & -- & 0.667 & 0.831 & 0.463 & \textbf{0.363} & -- & 0.363 & 0.508 & 0.448 & \textbf{0.102} \\
\bottomrule[0.8pt]
\end{tabular*}

\vspace{2pt}
{\raggedright\fontsize{5.4}{6.4}\selectfont RULER: MK1--3 niah\_multikey\_1--3, MQ niah\_multiquery, MV niah\_multivalue, S3 niah\_single\_3, VT variable tracking. LongBench categories: single-document QA, multi-document QA, summarization, few-shot, synthetic, code.\par}
\end{table*}

\subsection{Setup}

\textbf{Models and deployment.}
We evaluate Qwen3-4B and Qwen3-8B~\cite{yang2025qwen3}, Llama-3.2-3B~\cite{grattafiori2024llama3,meta2024llama32} and Gemma-3-4B~\cite{gemma2025gemma3}: the instruct models on the downstream tasks and, for Llama and Gemma, the base models for PPL.
All linear layers run in W4A4 NVFP4, queries in FP8 and values in NVFP4; the configurations differ only in the keys.
Accuracy is evaluated in software on four NVIDIA H100 GPUs; the software simulates the number formats of the accelerator for the weights, activations, queries, keys and values as well as the CarrierKV algorithm.
The software evaluation is aligned with the hardware.
On RULER and LongBench, Qwen3 runs in non-thinking mode.

\textbf{Benchmarks.}
RULER~\cite{hsieh2024ruler} runs with its official code and data generation on seven KV-sensitive subtasks of 13 at 32K, with 100 samples each.
LongBench v1~\cite{bai2024longbench} runs all 21 tasks with the official prompts, truncation and metrics.
PPL uses non-overlapping 4K- and 32K-token PG19 test windows~\cite{rae2020compressive} (2383 and 255 for Qwen3).

\textbf{Baselines.}
Besides NVFP4, we compare with two 4-bit key treatments: centering at the same cache size and the key format of an edge LLM accelerator.
SageAttention~\cite{sageattention2025} subtracts the mean key after RoPE; we take the mean of up to 512 prompt tokens after the first.
P3-LLM~\cite{p3llm2026} scales each key channel after RoPE and stores INT4 in groups of 128, as its paper does for Llama-3.

\textbf{Hardware.}
We implement the CarrierKV accelerator and the NVFP4 baseline accelerator, with $G = 4$ query heads per KV head as in Qwen3-4B and Qwen3-8B, in a 22\,nm FDSOI process at 600\,MHz with the same flow, measure their gate-level power on Qwen3-4B data and take the SRAM macros from the memory compiler.
The latency and energy per decode token, and per reasoning request with a 1K-token prompt and 20K generated tokens (about the mean output length on AIME, HMMT25 and LiveCodeBench~\cite{jain2025livecodebench}), come from our performance model, which combines SCALE-Sim v3~\cite{raj2025scalesim3} for the GEMM core with Ramulator 2.1~\cite{luo2026ramulator21} for the LPDDR5-6400 x64 DRAM, whose energy follows the device datasheet currents.
The NVFP4 baseline has the same GEMM core, SRAM and attention engine without the parts CarrierKV adds.

\subsection{Accuracy}

\begin{figure}[b]
\centering
\includegraphics[width=\columnwidth]{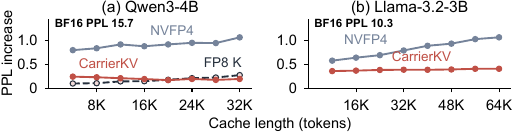}
\caption{PPL increase over the BF16 KV cache by cache length on PG19: (a) 32K windows in 4K segments; (b) 64K windows in 8K segments.}
\label{fig:cachelen}
\end{figure}

\textbf{Loss over the cache length.}
\Cref{fig:cachelen} splits the PPL increase by cache length.
On Qwen3-4B, the increase grows with NVFP4 keys from 0.80 in the first 4K to 1.07 in the last, and with FP8 keys from 0.10 to 0.27, while CarrierKV stays between 0.17 and 0.24 and is below FP8 keys in each of the last four segments (0.19 against 0.27 in the last; Qwen3-8B: 0.09 against 0.12).
On 64K windows of Llama-3.2-3B, the increase per 8K grows from 0.58 to 1.07 with NVFP4 and from 0.36 to 0.41 with CarrierKV; FP8 keys stay lower on Llama, from 0.10 to 0.24.

\textbf{Downstream tasks and PPL.}
\Cref{tab:accuracy} lists every RULER subtask and LongBench category.
Against NVFP4, CarrierKV raises the 32K RULER average on all four models and lowers the PPL increase at 4K and 32K on all four, at 32K by 1.8 to 5.9 times.
The gain is largest where NVFP4 degrades most: on Qwen3-4B, RULER rises from 79.4 to 91.1 (reference 92.6), and its hardest multi-key subtask, MK3, from 43 to 79 (reference 87).
Against the two baselines, CarrierKV has the lowest 32K PPL increase and the highest 32K RULER and LongBench averages on all four models; on Qwen3-4B, its RULER average is 91.1 against 86.9 with SageAttention and 75.1 with P3-LLM.
On Llama-3.2-3B, P3-LLM has a lower 4K PPL increase (0.348 against 0.349).

\begin{table}[t]
\centering
\caption{Qwen3-8B, thinking mode: accuracy (\%), mean $\pm$ std over runs.}\label{tab:reason}
{\small\setlength{\tabcolsep}{1pt}\renewcommand{\arraystretch}{1.0}%
\setlength{\aboverulesep}{0.2ex}\setlength{\belowrulesep}{0.2ex}%
\newcommand{\sd}[1]{{\footnotesize$\pm$#1}}%
\begin{tabular*}{\columnwidth}{@{\extracolsep{\fill}}lcccccc@{}}
\toprule[0.5pt]
 & AIME24+25 & HMMT25 & LCB & GPQA & Olymp. & Avg. \\
\midrule[0.4pt]
BF16 & 65.2\sd{4.3} & 37.1\sd{5.5} & 42.7\sd{1.5} & 52.9\sd{5.1} & 74.9 & 54.6 \\
NVFP4 & 58.5\sd{3.3} & 32.1\sd{6.7} & 41.5\sd{0.4} & 51.3\sd{1.6} & 72.0 & 51.1 \\
SageAttention & 57.1\sd{3.8} & 32.9\sd{5.8} & 42.0\sd{2.0} & 50.5\sd{1.3} & 73.1 & 51.1 \\
CarrierKV & \textbf{60.8}\sd{3.2} & \textbf{32.9}\sd{4.5} & \textbf{43.0}\sd{1.9} & \textbf{52.9}\sd{3.7} & \textbf{73.4} & \textbf{52.6} \\
\bottomrule[0.5pt]
\end{tabular*}}
\par\vspace{0.3ex}
{\raggedright\fontsize{5.4}{6.4}\selectfont LCB: LiveCodeBench v6; GPQA: GPQA-Diamond; Olymp.: OlympiadBench. Runs: 8, 8, 3, 3. Bold: best 4-bit.\par}
\end{table}

\textbf{Reasoning.}
With thinking on, Qwen3-8B generates up to 32K tokens from short prompts, so its cache holds mostly generated keys.
\Cref{tab:reason} gives its accuracy on AIME24, AIME25, HMMT25, LiveCodeBench v6~\cite{jain2025livecodebench}, GPQA-Diamond~\cite{rein2024gpqa} and OlympiadBench~\cite{he2024olympiadbench}.
CarrierKV has the highest average of the 4-bit configurations, 52.6 against 51.1 with NVFP4 and with SageAttention (BF16: 54.6), and leads or ties them on every benchmark.

\subsection{Hardware}

\begin{figure}[t]
\centering
\includegraphics[width=\columnwidth]{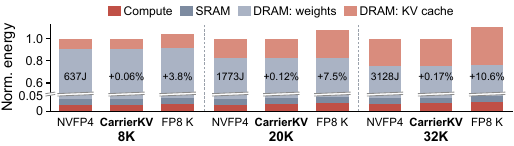}
\caption{Energy of a reasoning request on Qwen3-4B with a 1K-token prompt and 8K, 20K or 32K generated tokens, normalized to NVFP4.}
\label{fig:energytoken}
\end{figure}

\textbf{Area and energy.}
\Cref{tab:ppa} lists the area and the energy per generated token of a reasoning request on Qwen3-4B, prefill included.
CarrierKV's hardware increases the area of the NVFP4 accelerator by 2.9\% and this energy by 0.12\%, because DRAM accounts for 96\% of it; on Qwen3-8B, the energy increase is 0.07\%, and on both models the latency grows by less than 0.01\%.
From 8K to 32K generated tokens, the increase grows from 0.06\% to 0.17\% on Qwen3-4B (\cref{fig:energytoken}) and from 0.04\% to 0.10\% on Qwen3-8B, while FP8 keys use 3.8\% to 10.6\% more energy and 3.7\% to 10.2\% more time than NVFP4 on Qwen3-4B.
On both models, CarrierKV reaches a PPL increase (\cref{fig:cachelen}) and a 32K RULER average similar to those of FP8 keys (91.1 and 97.4 against 89.3 and 97.1) and, on Qwen3-4B, is 1.16 times faster with 13.7\% less energy per decode step at a 32K-token cache, and 1.28 times faster with 22.2\% less at 128K.

\begin{table}[t]
\centering
\caption{Area and energy per generated token (Qwen3-4B, 1K-token prompt, 20K generated tokens; 22\,nm FDSOI, 0.8\,V, 600\,MHz).}\label{tab:ppa}
{\footnotesize\setlength{\tabcolsep}{2pt}\renewcommand{\arraystretch}{0.9}%
\setlength{\aboverulesep}{0.2ex}\setlength{\belowrulesep}{0.2ex}%
\begin{tabular*}{\columnwidth}{@{\extracolsep{\fill}}lrr@{}}
\toprule[0.5pt]
Component & Area (mm$^2$) & mJ/token \\
\midrule[0.4pt]
NVFP4 accelerator & 4.945 & 86.595 \\
\midrule[0.4pt]
CarrierKV accelerator & \textbf{5.087} {\scriptsize(+2.9\%)} & \textbf{86.700} {\scriptsize(+0.12\%)} \\
\quad GEMM core & 1.467 & 1.041 \\
\quad SRAM & 2.925 & 1.656 \\
\quad Attention engine & 0.405 & 0.723 \\
\qquad Score unit & 0.182 & 0.330 \\
\qquad\quad QK$^\top$ unit & 0.086 & 0.231 \\
\qquad\quad Carrier unit & 0.096 & 0.099 \\
\qquad PV unit, softmax & 0.164 & 0.393 \\
\qquad KV encoder & 0.059 & $<$0.001 \\
\quad Prefill units & 0.078 & $<$0.001 \\
\quad Vector unit & 0.136 & 0.047 \\
\quad RISC-V core, DMA & 0.076 & 0.162 \\
\quad LPDDR5 DRAM & -- & 83.071 \\
\bottomrule[0.5pt]
\end{tabular*}}
\end{table}

\textbf{Against a cache before RoPE.}
A cache before RoPE, as in KVQuant~\cite{hooper2024kvquant}, must decode and rotate every cached key on each read.
At the same throughput, its BF16 dot-product unit with FP8 queries and RoPE add at least 166\,pJ per key to the NVFP4 score unit, 5.2 times the 32\,pJ that CarrierKV adds.

\section{Conclusion}\label{sec:conclusion}

Most of the energy in the cached keys is a positional carrier, which can be computed from their positions, so CarrierKV removes it before RoPE with one vector per KV head and regenerates its score term on chip with a positional phase engine.
At the same cache size as NVFP4, CarrierKV raises 32K RULER on Qwen3-4B from 79.4 to 91.1 and lowers the 32K PPL increase of NVFP4 by 1.8 to 5.9 times on four models, while its hardware increases the area of the NVFP4 accelerator by 2.9\% and the energy of a 20K-token reasoning request on Qwen3-4B by 0.12\%.

\section*{Acknowledgment}
The authors thank Google Cloud for providing compute resources, Google Research for supporting this work, and GlobalFoundries and its University Partnership Program for technology access.

\bibliographystyle{IEEEtran}
\bibliography{refs}

\end{document}